\documentclass[aps,pre,reprint,superscriptaddress,nofootinbib]{revtex4-1}
\usepackage{graphicx}
\usepackage{amsmath}
\usepackage{amsfonts}
\usepackage{hyperref}
\begin{document}

\title{Strain fluctuations and elastic moduli in disordered solids}

\author{Daniel M. Sussman}
\email[]{dsussman@sas.upenn.edu}
\thanks{DMS and SSS contributed equally to this work}
\affiliation{Department of Physics and Astronomy, University of Pennsylvania, 209 South 33rd Street, Philadelphia, Pennsylvania 19104, USA}
\author{Samuel S. Schoenholz}
\affiliation{Department of Physics and Astronomy, University of Pennsylvania, 209 South 33rd Street, Philadelphia, Pennsylvania 19104, USA}
\email[]{schsam@sas.upenn.edu}
\author{Ye Xu}
\affiliation{Department of Physics and Astronomy, University of Pennsylvania, 209 South 33rd Street, Philadelphia, Pennsylvania 19104, USA}
\affiliation{Complex Assemblies of Soft Matter, CNRS-Rhodia-UPenn UMI 3254, Bristol, Pennsylvania 19007, USA}
\author{Tim Still}
\affiliation{Department of Physics and Astronomy, University of Pennsylvania, 209 South 33rd Street, Philadelphia, Pennsylvania 19104, USA}
\author{A. G. Yodh}
\affiliation{Department of Physics and Astronomy, University of Pennsylvania, 209 South 33rd Street, Philadelphia, Pennsylvania 19104, USA}
\author{Andrea J. Liu}
\affiliation{Department of Physics and Astronomy, University of Pennsylvania, 209 South 33rd Street, Philadelphia, Pennsylvania 19104, USA}
\date{\today}

\begin{abstract}
Recently there has been a surge in interest in using video-microscopy techniques to infer the local mechanical properties of disordered solids. One common approach is to minimize the difference between particle vibrational displacements in a local coarse-graining volume and the displacements that would result from a best-fit affine deformation. Effective moduli are then be inferred under the assumption that the components of this best-fit affine deformation tensor have a Boltzmann distribution. In this paper, we combine theoretical arguments with experimental and simulation data to demonstrate that the above does not reveal information about the true elastic moduli of jammed packings and colloidal glasses.
\end{abstract}

\maketitle

\section{Introduction}
Characterizing the elasticity of soft disordered materials is challenging, in part because these systems often lie at the boundary of where classical theories of elasticity are applicable \cite{landau1986}. In jammed harmonic sphere packings, for example, a length scale below which continuum elasticity breaks down has been explicitly identified, and this length diverges upon the approach to the jamming transition \cite{Lerner2014}. Another complication derives from the presence of nonaffine distortions that disordered materials experience in response to imposed deformations. It has been argued that these nonaffinities arise from spatial fluctuations of local elastic moduli \cite{DiDonna2005} and finite-temperature effects \cite{Sengupta2008, Ganguly2013}.

In the last decade there has been a surge of experiments and simulations that aim to calculate elastic constants from particle-level fluctuations \cite{Zahn2003, Schall2006, Schall2007, Zhang2009, Schall2011, Schall2012}. One common approach focuses on thermally induced microscopic strain fluctuations \cite{Sengupta2000,Sengupta2008, Sengupta2010,Schall2007}. In these studies, the fluctuations of a locally-defined strain field are aggregated over time to arrive at a distribution of strains at each point. It is assumed that these strains are drawn from a Boltzmann distribution whose weight defines a local elastic modulus. Global elastic properties of the system are then computed by averaging or coarse graining these local moduli. In the case of crystalline systems \cite{Sengupta2000, Sengupta2010} it is relatively straightforward to construct a local strain field on a per-particle basis by appealing to an undeformed lattice. In the case of amorphous systems, however, such an identification is no longer possible. Instead, it has been suggested that one should construct a coarse grained strain field by computing the best affine approximation to the collective motion that some neighborhood of particles undergo \cite{Schall2007}. 

In this paper we argue that thermally induced particle motion in amorphous solids does not permit the use of local affine strain distributions (computed as best-fit affine transformations) to compute elastic moduli. To demonstrate this concept we consider two systems: a simulation of harmonic disks in two- and three-dimensions and a quasi-two-dimensional experimental colloidal system, studied and described in detail in Ref. \cite{Still2014}. We rely, in particular, on simulations of harmonic disks since their elastic moduli have been very well characterized by independent measurements and theoretical analyses. Moreover, these systems can be simulated at arbitrarily low temperatures to ensure that we are truly in the regime where linear response is valid. One quantification of the degree of structural correlations captured by the best-fit affine transformation, $\Lambda$, are their dependence on the coarse-graining scale $L$. If $\Lambda$ is to relate to linear elasticity then we should expect $\text{var}[\Lambda_{\alpha\beta}]\sim L^{-d}$ as expressed by~\cite{Schall2007}, where $\alpha$ and $\beta$ specify the indices of the deformation tensor. We will use this simple scaling relation as a benchmark throughout our analysis, and we will show that the structural correlations captured by $\Lambda_{\alpha\beta}$ are systematically too weak to relate to the elastic moduli.

We wish to emphasize that defining best-fit local strains in actively deformed materials can, in conjunction with knowledge of the local stresses, still lead to meaningful information about the elasticity of disordered materials \cite{Barrat2009}. Additionally, tracking particle positional information over a long time can be effectively used to estimate the covariance matrix of the system \cite{Still2014}. However, we argue that if there are only thermally induced fluctuations, then the distribution of best-fit strains does not contain information about the system elasticity.

In Sec. \ref{sec:formal} we introduce the formalism commonly used to extract affine strains from thermal fluctuations, and we highlight some problems with interpreting exponential fits of the associated distribution to obtain elastic moduli. In Sec. \ref{sec:sdt} we study the simulations in a regime where the correlations between particle positions and displacements are relatively small, and we show that a simple statistical model completely describes the distributions of local strains. The statistical model accurately predicts the distriubtions for for all temperatures, coarse-graining sizes, and pressures, and we show that these measurements are emphatically \emph{not} connected to the elastic moduli of the systems. In Sec. \ref{sec:ldt} we discuss simulation and experimental measurements in a regime with increasingly large particle position and displacement correlations. While the correlations that enter the calculation of the best-fit affine deformation tensor in this regime might be expected to enable one to deduce the elastic moduli, we again find that this is not the case. We discuss these results and their consequences for interpreting experimental data in Sec. \ref{sec:disc}. The Appendices present a reformulation of the quantity $D^2_{min}$ used to define $\Lambda_{\alpha\beta}$ (Appendix \ref{sec:formalism}), the statistical model used to understand the data in Sec. \ref{sec:sdt} (Appendix \ref{sec:statmodel}), and the details of the simulations and experimental systems studied (Appendix \ref{sec:simdetail}).

\section{Identifying local strains from thermal fluctuations
\label{sec:formal}}

We begin by introducing the formalism that has most commonly been used to extract affine strains from thermal fluctuations. As mentioned above, the most-studied practical solution to the problem of nonaffinities is to find the best-fit affine transformation of local particle positions at time $t-\Delta t$ onto particle positions at time $t$, and then study distributions associated with this affine component of particle motion. This scheme is commonly done using a measure of nonaffinity, $D^2_{min}$, originally introduced by Falk and Langer \cite{Falk1998}. One version of this calculation considers a local square or cubic coarse graining volume of side length $L$ centered at point $\vec{R}$. The motion of the particles $j$ in that local volume are tracked between times $t-\Delta t$, and $t$ and then one computes the deviation of their displacements from those described by a best-fit affine transformation over that time window \cite{Falk1998}. Explicitly,
\begin{align}
D^2(t,\Delta t) &= \sum_j \sum_\alpha \bigg( r_j^\alpha(t)-R^\alpha\nonumber\\
&\hspace{2pc} - \sum_\beta (\delta_{\alpha\beta} + \Lambda_{\alpha\beta})\left[ r_j^\beta (t-\Delta t)-R^\beta \right]  \bigg)^2
\end{align}
Here the Greek indices run over the Cartesian coordinates, $\vec{r}_j(t)$ is the position of particle $j$ at time $t$, $\Lambda$ is an affine transformation tensor, and $\delta_{\alpha\beta}$ is the Kronecker delta. This quantity is then minimized over all possible affine transformation tensors, $\Lambda_{\alpha\beta}$:
\begin{equation}
D^2_{min}(t,\Delta t) = \min_{\Lambda_{\alpha\beta}}D^2(t,\Delta t).
\end{equation}
Solving for the minimizing affine transformation is straightforward. Defining
\begin{equation}
X_{\alpha\beta} = \sum_j ( r_j^\alpha(t) - R^\alpha ) \times ( r_j^\beta(t-\Delta t) - R^\beta ),
\end{equation}
\begin{equation}
Y_{\alpha\beta} = \sum_j ( r_j^\alpha(t-\Delta t) - R^\alpha ) \times ( r_j^\beta(t-\Delta t) - R^\beta ),
\end{equation}
the best-fit tensor can be written as
\begin{equation}
\Lambda_{\alpha\beta} = \sum_\gamma X_{\alpha\gamma} Y_{\beta\gamma}^{-1}-\delta_{\alpha\beta}.
\end{equation}
The standard approach~\cite{Schall2007} has then been to assume that these strains are drawn from a Boltzmann distribution whose energy is given by the elastic energy $E/\mu = \Lambda_{\alpha\beta}^2L^d$ where $\mu$ is the elastic constant. If this assumption holds, then the probability distribution of the squared strain components is given by $P(\Lambda_{\alpha\beta}^2) \sim \exp(-\mu\Lambda_{\alpha\beta}^2L^d/kT)$ and the modulus can be extracted by fitting the logarithm of the probability distribution to a straight line.

\begin{figure} 
\centering{\includegraphics[width=0.8\linewidth]{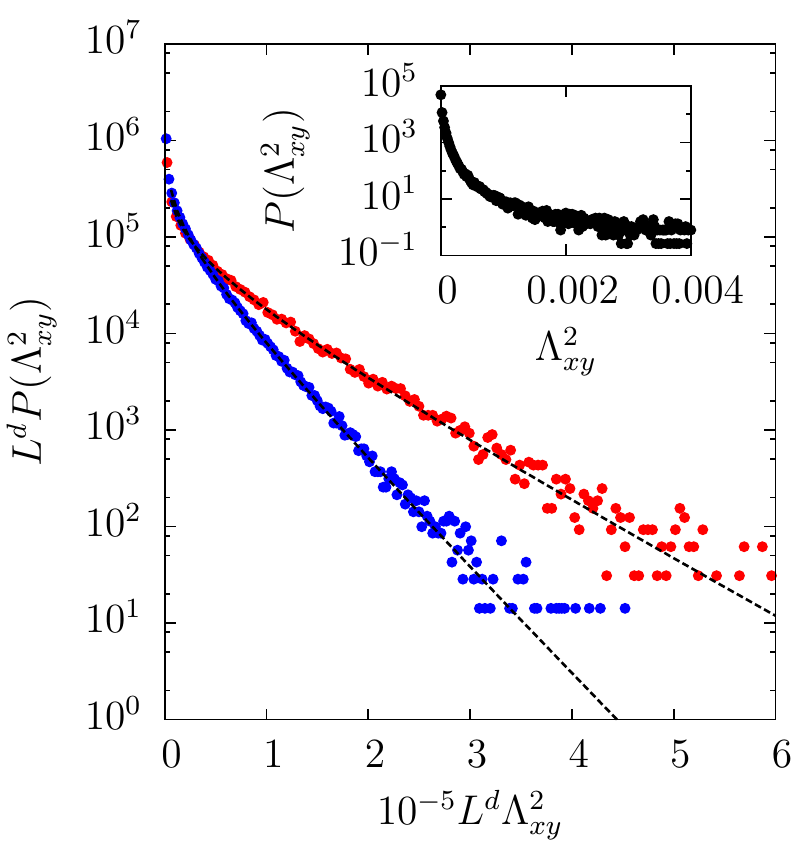} }
\caption{\label{fig:exy2}
(Color online) Probability distribution of $L^d \Lambda_{xy}^2$ for simulations in $d=2$ (upper red circles) and $d=3$ (lower blue circles) of harmonic repulsive spheres at $T=10^{-5}$. Curves are best-fit $\chi^2$ distributions. Inset: Probability distribution of the unscaled $\Lambda_{xy}^2$ for the experimental colloidal system.}
\end{figure}

To test this basic claim we plot, in Fig. \ref{fig:exy2}, the distributions of $\Lambda_{\alpha\beta}^2$ as measured for small $\Delta t$ for both 2D and 3D harmonically repulsive disks, as well as for the colloidal system at longer $\Delta t$. We find that this distribution has a pronounced curvature on a log-linear scale, suggesting that a simple exponential decay is a poor description of the distribution. A similar curvature in this distribution was reported by Rahmani et al., where it was interpreted to result from a heterogeneous distribution of local moduli \cite{Schall2014}. However, as shown in the figure, we find that the simulation distributions are accurately described by a simple $\chi^2$ distribution coming from the square of a single Gaussian random variable. Indeed, Ganguly et al. showed that a perfect hexagonal lattice at low temperatures has a Gaussian distribution of $\Lambda_{xy}$ \cite{Ganguly2013}, implying that there, too, the distribution of $\Lambda_{xy}^2$ would  take a $\chi^2$ form and not appear as a straight line on a log-linear plot. Given that this qualitative signature of a heterogeneous distribution of elastic moduli can be completely reproduced in the context of a simple statistical model with a single underlying Gaussian distribution (as discussed in more detail below), our results for disordered solids reinforce the message that it is hazardous to fit exponential decays to these distributions and then interpret the result as elastic moduli.

We note that an alternate definition of the best-fit affine deformation tensor that is commonly used, and is due to Cundall \cite{Cundall1979}, first subtracts rigid body displacements of the local clusterÕs center of mass before computing the affine distortion. Since amorphous materials can have large low-energy fluctuations compared to crystalline systems, these center of mass displacements could, in principle, be important in our analysis, particularly for relatively small coarse graining lengths. However, we have carefully checked that the conclusions in our paper are insensitive to this definitional choice.

In the following we will systematically study the distributions of $\Lambda_{\alpha\beta}$ from our simulations as a function of initial system pressure $p$, the time window $\Delta t$, coarse-graining scale $L$, and temperature $T$. We will supplement this with experimental data for a 2D colloidal sample. By varying the $\Delta t$ at which we compute the best-fit affine deformation tensor between the ballistic regime and the cage regime we can systematically tune the amount of correlation between displacements of a particle and its local environment. In no regime do we find a connection between measured distributions of affine strains and elastic moduli.

\section{Strain measurements for small $\Delta t$
\label{sec:sdt}}

In the ballistic regime of small $\Delta t$ (where time is measured in units of $\tau$, the Lennard-Jones-like time unit of our simulation) there are only relatively small correlations between the frame-to-frame displacements of particles and their initial positions. As such, we expect that the measurement of locally coarse-grained $\Lambda_{\alpha\beta}$ can be understood as the result of measuring single-point particle fluctuations and then averaging over a locally amorphous environment. Although our experimental data is well out of this regime, it is easily probed in our simulations. Clearly, in this regime the system lacks the particle-displacement correlations necessary to be described as a solid. Nevertheless, our exploration of this regime enables us to unambiguously identify an existing problem in how strain distributions have been analyzed in amorphous solids \cite{Schall2007, Schall2014}. It also enables us to set up a convenient metric for how the strain variances must vary with coarse graining scale in order to be interpreted as moduli.

As we show in Appendix \ref{sec:formalism}, the calculation of the best-fit affine deformation tensor in a local coarse-graining volume can be usefully formulated as
\begin{equation}\label{eq:lambdareform}
\Lambda_{\alpha\beta} = \sum_j \Delta_{j\alpha}\left(\sum_\gamma A_\gamma r_{j\gamma}\right),
\end{equation}
where the first sum is over all particles $j$ in the local coarse-graining volume, $\Delta_{j\alpha}$ is the frame-to-frame displacement of particle $j$ in the $\alpha$ direction, the $A_\gamma$ are quantities related to the initial positions of all particles in the local volume, and $r_{j\gamma}$ is the $\gamma$ component of the position of particle $j$ at time $t-\Delta t$. In the limit of small correlations between displacements and local structure, then, $\Lambda_{\alpha\beta}$ can be approximated as a sum of random variables, where $\Delta_{j\alpha}$ is drawn from a Gaussian distribution whose width, $\sigma_\Delta$, is set by the density and temperature of the system and which is uncorrelated with the structural parameters $A_\gamma$. 

In Appendix \ref{sec:statmodel} we combine this idea with the simplest possible model of the spatial structural parameters $A_\gamma$, treating the positions of particles within the local coarse-graining volume as being uniformly distributed with no excluded volume. By the central limit theorem this simple statistical model predicts that $\Lambda_{\alpha\beta}$ will have an approximately Gaussian distribution. Given a measurement of $\sigma_\Delta$ and the particle number density, Appendix \ref{sec:statmodel} provides a prediction for its variance.

\begin{figure*}[ht!]
\centering{\includegraphics[height=0.3\linewidth]{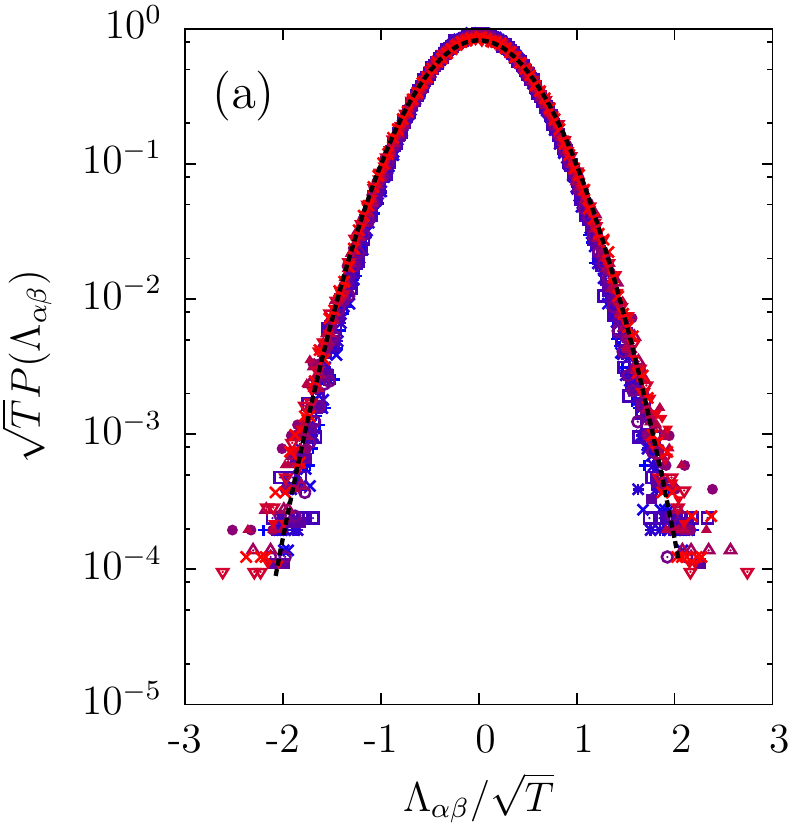}\hspace{1pc}\includegraphics[height=0.3\linewidth]{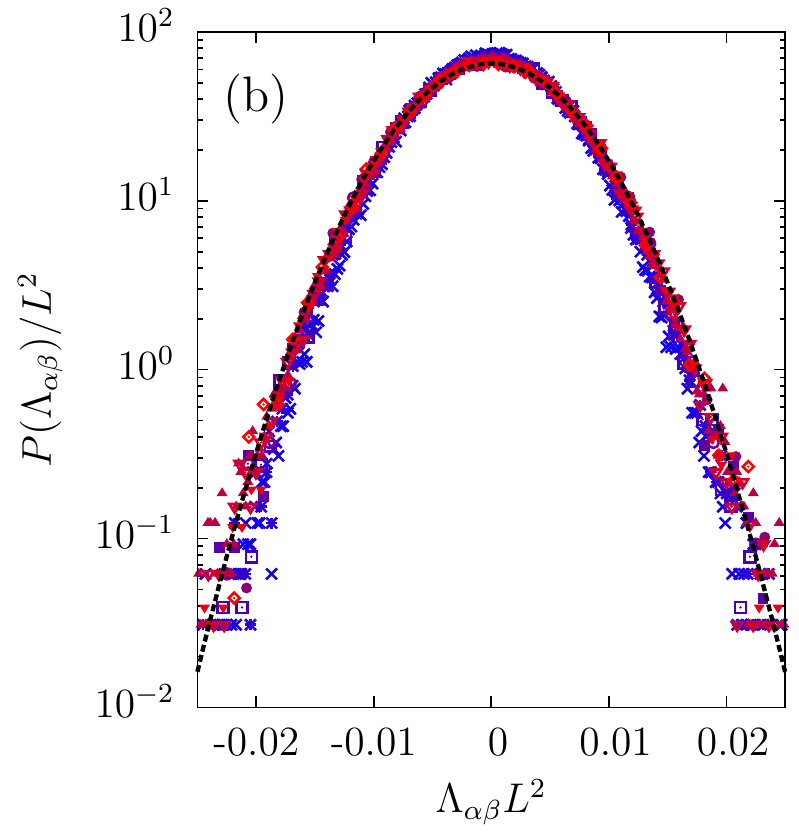}}\hspace{1pc}
\centering{\includegraphics[height=0.3\linewidth]{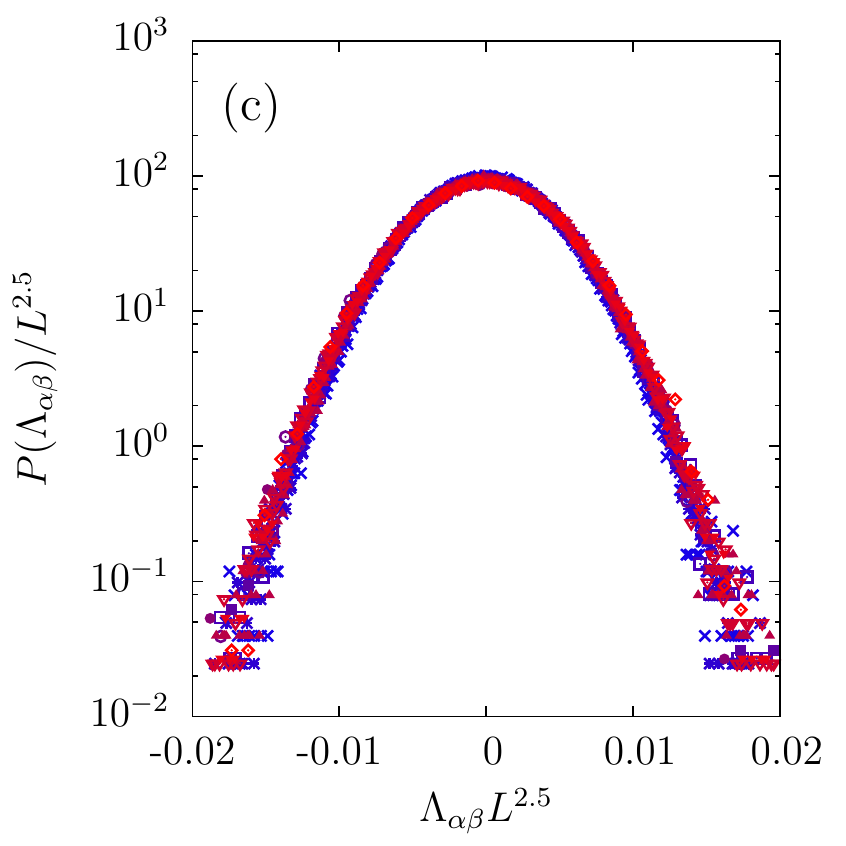}}
\caption{\label{fig:collapse_sdt} (Color online) Collapsed probability distributions of $P(\Lambda_{xx})$ (blue points) and $P(\Lambda_{xy})$ (red points) as computed for $\Delta t = 2\tau$. (A) Scaling collapse of $P(\Lambda_{\alpha\beta})$ for 2D simulations with $\sqrt{T}$ for coarse-graining scale $L=2$, $p=10^{-2}$ and temperatures of $T =10^{-5},\ 2\times 10^{-5},\ 4\times 10^{-5},\ 6\times 10^{-5},\ 8\times 10^{-5},\ 10^{-4}$. Dashed line is the prediction from Appendix \ref{sec:statmodel}. (B) Scaling collapse of $P(\Lambda_{\alpha\beta})$ for 2D simulations with $L^2$ for $T=10^{-5}$, $p=10^{-2}$, and $L = 2,\ 4,\ 6,\ 8,\ 10,\ 12,\ 14$. Dashed line is the prediction from Appendix \ref{sec:statmodel} for the $L=6$ data set. (C) Scaling collapse of $P(\Lambda_{\alpha\beta})$ for 3D simulations with $L^{2.5}$ for $T=10^{-5}$, $p=10^{-2}$, and $L = 2,\ 4,\ 6,\ 8,\ 10,\ 12,\ 14$.}
\end{figure*}

Despite the naivete of this model, we find that the distributions of $\Lambda_{\alpha\beta}$ when measured with a small $\Delta t$ are remarkably well described by sums of Gaussian random variables multiplied by uncorrelated, ideal-gas-like structural parameters. Figure \ref{fig:collapse_sdt} demonstrates this, showing that the distributions of $\Lambda_{xx}$ and $\Lambda_{xy}$ collapse when scaled by the appropriate powers of temperature and coarse-graining scale predicted by the model. Additionally, as expected by the model, the distributions for every component of $\Lambda_{\alpha\beta}$ is nearly identical. Even more remarkably, the simple statistical model predicts the variance of the observed Gaussian distributions to within $10\%$.

\begin{figure} 
\centering{\includegraphics[width=0.8\linewidth]{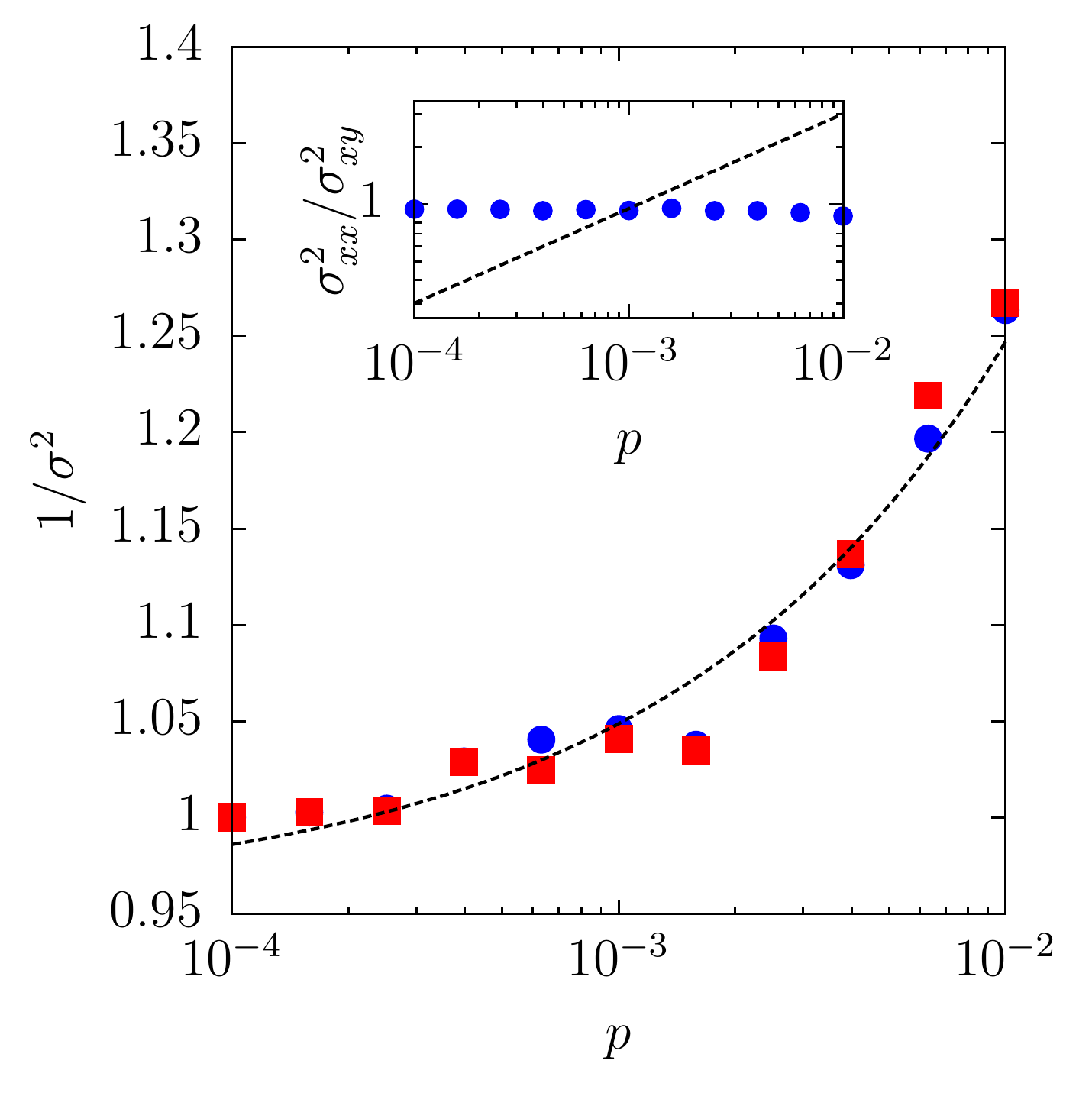}}
\caption{\label{fig:moduli_sdt}
(Color online) Inverse variance of $\Lambda_{xx}$ (blue circles) and individual particle displacement magnitude (red squares) for 2D simulations with $T=10^{-5}$ and $L=4$ as a function of pressure, normalized by the value at $p=10^{-4}$. Dashed line is the predicted scaling of the bulk modulus with pressure. Inset: Ratio of the variance of the diagonal to off-diagonal component of $\Lambda_{\alpha\beta}$ as a function of pressure. The corresponding ratio of moduli scales as $G/B\sim \sqrt{p}$ (dashed line).}
\end{figure}

Given that the variances are so well-described by a statistical model with no positional correlations and no correlations between particle positions and displacements, any effort to extract elastic moduli from this measurement is doomed to failure. As a simple demonstration of this failure, we follow Ganguly et al. and interpret $\textrm{var}\left[ \Lambda_{xx}+\Lambda_{yy} \right]$ as the bulk compliance and $\textrm{var}\left[ \Lambda_{xy}+\Lambda_{yx} \right]$ as the shear compliance \cite{Ganguly2013} (given that $P(\Lambda_{\alpha\beta}^2)$ is so well-described by a $\chi^2$ distribution it makes little sense to fit a straight line to it).  Explicitly, as a function of coarse-graining volume $L$ the relationship between the bulk and shear modulus and local strain fluctuations is given by
\begin{eqnarray}
\textrm{var}\left[ \Lambda_{xx}+\Lambda_{yy} \right]&=&\frac{k_B T}{L^2}\left(B(L)+G(L) \right)^{-1} \\
\textrm{var}\left[ \Lambda_{xy}+\Lambda_{yx} \right]&=&\frac{k_B T}{4 L^2}\left(G(L) \right)^{-1}.
\end{eqnarray}
We then plot the moduli -- the inverse compliances -- in Fig. \ref{fig:moduli_sdt} as a function of pressure for our simulated systems. Notably, the ratio of the bulk to the shear modulus is constant, whereas it is known that these jammed packings have a ratio that scales with the square root of the pressure, $G/B \sim \sqrt{p}$ ~\cite{Liu2010}. Both the bulk and shear moduli scale with the true bulk modulus of the system (the dashed line in the figure); the overall scale of fluctuations, $\sigma_\Delta$, itself tracks the scaling of the bulk modulus. Thus, this measurement \emph{could} be used to extract the scaling of the bulk modulus with, e.g., pressure, but not its absolute magnitude -- of course, this scaling can be more more easily extracted by simply measuring the scale of the fluctuations directly.

The fact that in the small $\Delta t$ regime the $\Lambda_{\alpha\beta}$ are single-particle quantities whose variance scales as $L^{-4}$ in two dimensions and $L^{-5}$ in three dimensions is a useful reference. Recall that the energy proposed by Schall et al. scales as $E/\mu = \Lambda_{\alpha\beta}^2L^d$ and moreover that the mean of $\Lambda_{\alpha\beta}$ is zero. It follows that if $E/\mu$ is to be well-defined in the limit of large coarse graining sizes the variance of $\Lambda_{\alpha\beta}$ must scale as $L^{-d}$. Therefore in the regime of small $\Delta t$, where $\textrm{var}\left[ \Lambda_{\alpha\beta}\right]\sim L^{-d-2}$, the variance of $\Lambda_{\alpha\beta}$ cannot yield a well defined elastic modulus. Even though Fig. \ref{fig:moduli_sdt} already demonstrated this to be the case, the necessary condition that $\textrm{var}\left[ \Lambda_{\alpha\beta}\right]\sim L^{-d}$ in the range of $L$ considered will be useful to keep in mind when we continue to the cage regime where we can no longer rely on an analytic model.

\section{Measurements in the plateau regime of the mean-squared displacement \label{sec:ldt}}

The above section is a useful illustration that distributions of $\Lambda_{\alpha\beta}^2$ should not always be interpreted in terms of elastic moduli. However, it is perhaps not surprising that the relatively uncorrelated nature of particle positions and frame-to-frame displacements in the ballistic or crossover regimes of $\Delta t$ can be understood in terms of sums of Gaussian random variables and not in terms of elastic moduli. By increasing the $\Delta t$ we can continuously tune the degree of correlation between the particle displacements and the local structure, and one hypothesis is that it is precisely these correlations that allow one to probe the elastic moduli of the system in question.

\begin{figure} 
\centering{\hspace{0.5pc}\includegraphics[height=0.8\linewidth]{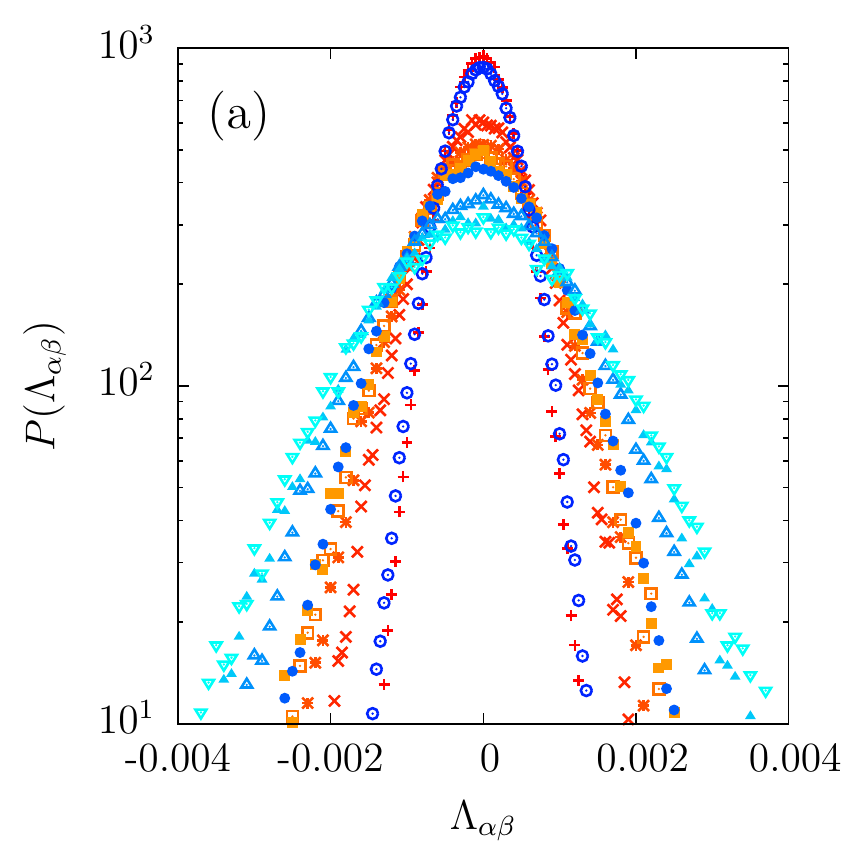} \includegraphics[height=0.8\linewidth]{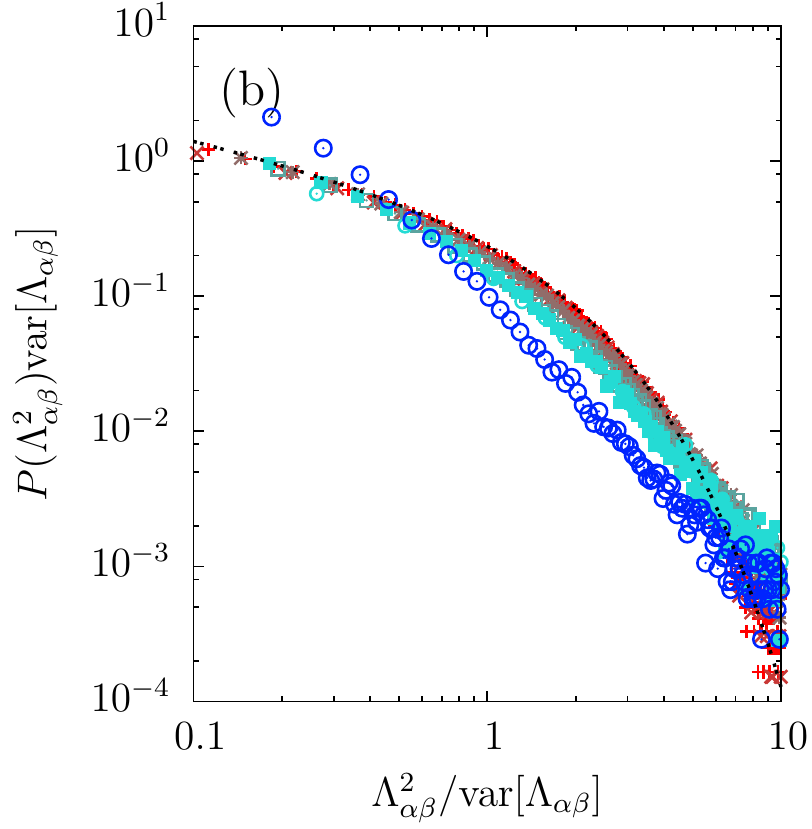}}
\caption{\label{fig:collapse_ldt}
(Color online) (A) $P(\Lambda_{xx})$ (red-to-orange color scale) and $P(\Lambda_{xy})$ (dark-to-light blue color scale) for 2D simulation data with $p=10^{-2}$, $T=10^{-5}$, $L=4$ for $\Delta t/\tau = 2,\ 10,\ 20,\ 50,\ 100,\ 200$. Data sets with smaller variance correspond to shorter $\Delta t$. (B) $P(\Lambda_{xy}^2)$ normalized by the variance for the above $\Delta t$ (red points correspond to the shortest $\Delta t$, light blue points to the longest $\Delta t$). Dark blue open circles are experimental data. The solid curve shows a unit-variance $\chi^2$ function, which matches the short-time data very well.}
\end{figure}

In Fig. \ref{fig:collapse_ldt} we show how the distributions $P(\Lambda_{\alpha\beta})$ change as $\Delta t$ is continuously increased from the ballistic regime to deep into the plateau regime. The most apparent change is that the variance of the diagonal and off-diagonal components of $\Lambda_{\alpha\beta}$ begin to separate. However, we also see that the distributions become increasingly non-Gaussian. Figure \ref{fig:collapse_ldt}B highlights the change in the tails of these distributions, showing that as $\Delta t$ is increased the distribution of $P(\Lambda_{\alpha\beta}^2)$ continuously shifts from being extremely well-characterized by a $\chi^2$ fit to one with an apparent power-law tail. We note that the experimental data in this plot is much deeper into the plateau regime of the mean-squared displacement than our simulation data. Nevertheless, the robust presence of a power-law tail in both the simulations and experiments further emphasizes the danger of fitting exponential decays to different parts of $P(\Lambda_{\alpha\beta}^2)$ distributions and of interpreting those fits as elastic moduli.

\begin{figure} 
\centering{
\hspace{0.5pc}\includegraphics[height=0.8\linewidth]{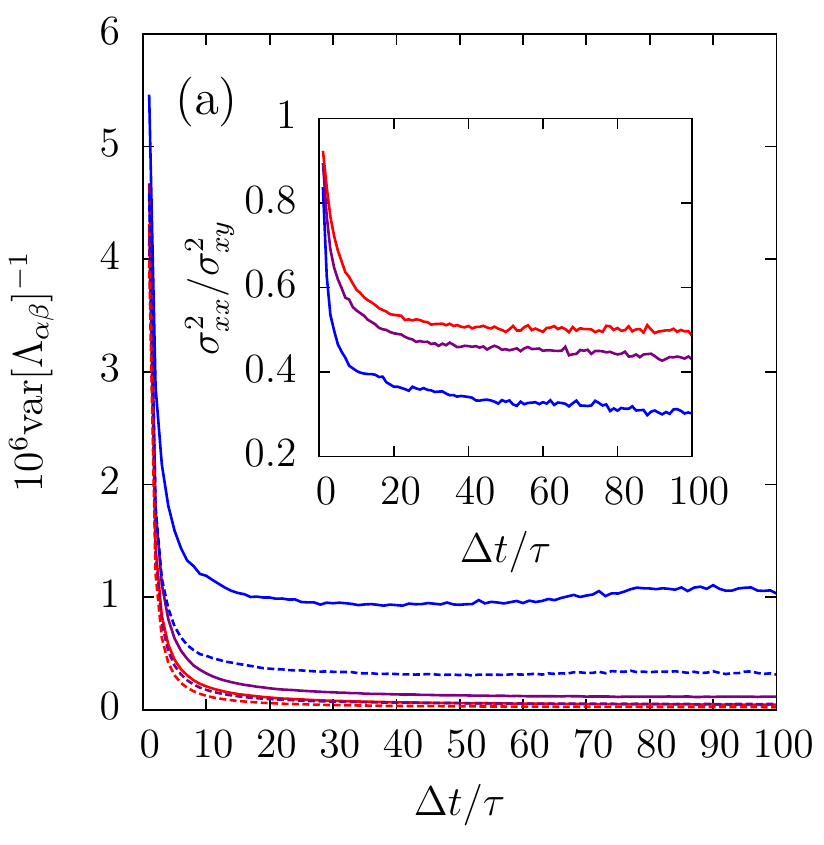} \includegraphics[height=0.8\linewidth]{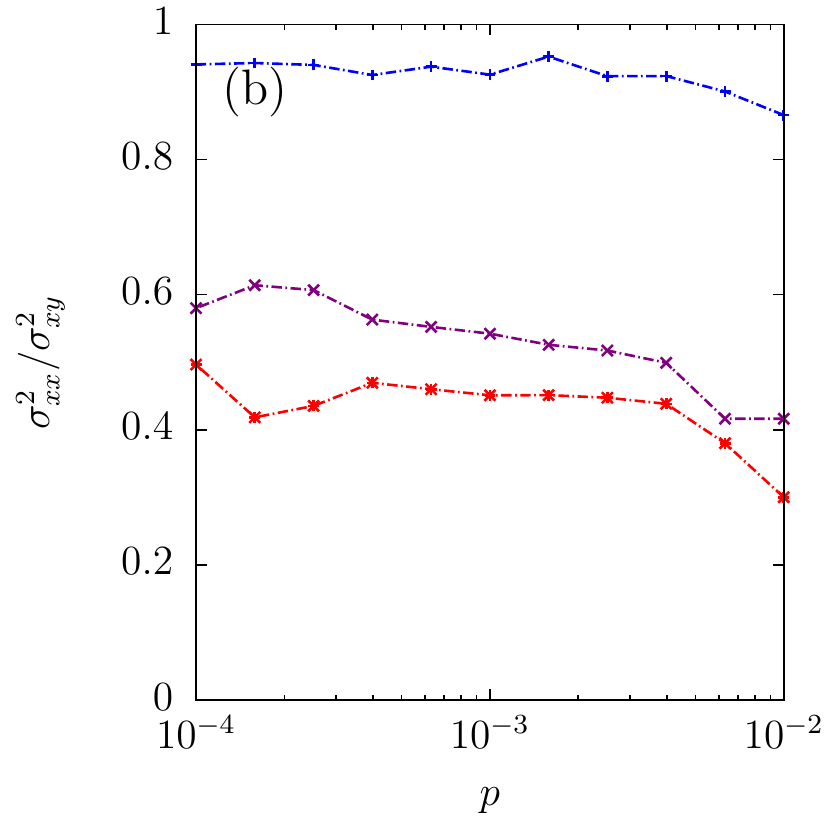}}
\caption{\label{fig:moduli_ldt}
(Color online) (A) Inverse variance of $P(\Lambda_{xx})$ as a function of $\Delta t$ for pressures between $p=10^{-2}$ (blue, upper curve) and $p=10^{-4}$ (red, lower curve). Inset: Ratio of the variance of the diagonal to the off-diagonal components of $\Lambda_{\alpha\beta}$ as a function of $\Delta t$ (B) Ratio of variance of diagonal and off-diagonal components of $\Lambda_{\alpha\beta}$ as a function of pressure for $\Delta t/\tau = 2,20,200$. The corresponding ratio of moduli scales as $G/B\sim \sqrt{p}$.}
\end{figure}

However, given that there are clearly some increasing correlations being picked up by the  $P(\Lambda_{\alpha\beta})$ distributions, one may wonder if the variances of these distributions are related to the elastic compliances. In Fig. \ref{fig:moduli_ldt} we plot inverse variances as a function of $\Delta t$ and $p$. Once again we see that the ratio of moduli estimated in this way is independent of pressure, in stark contrast to the true elastic constants of these systems. Thus, although there are additional correlations in these data sets, they are not straightforwardly connected to the appropriate integrals over the covariance matrix that would allow one to correctly extract elastic constants.

\begin{figure} 
\centering{
\includegraphics[width=0.8\linewidth]{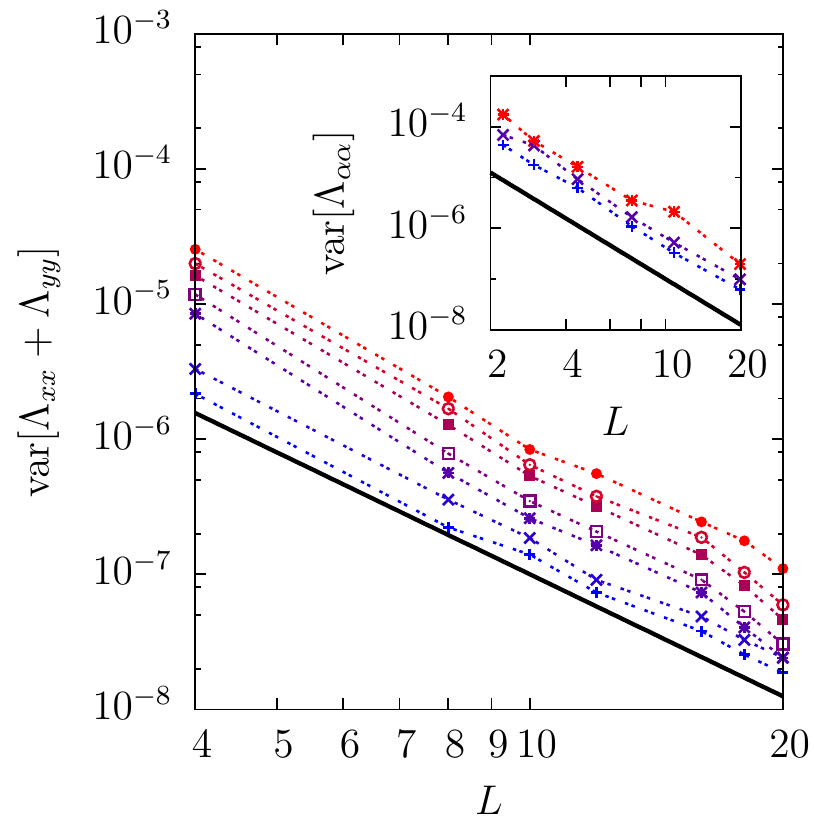} }
\caption{(Color online) \label{fig:moduli_ldt_L} $\text{var}[\Lambda_{xx}+\Lambda_{yy}]$, as a function of coarse graining size $L$ in the simulation of two-dimensional harmonic disks for pressures from $p = 10^{-2}$ (blue, bottom lines) to $p=10^{-3.4}$ (red, top lines). Overlaid in black is a line with slope $-3$. Here $\Lambda_{\alpha\beta}$ was calculated with $\Delta t = 5\times10^4$, deep into the caged regime. Inset: $\text{var}[\Lambda_{xx}+\Lambda_{yy}]$ as a function of coarse-graining scale (measured in microns) for area fractions $\phi = 0.8625, 0.8695, 0.8822$ (top to bottom).
}
\end{figure}

Finally, as in the case of our measurements at short $\Delta t$, we can attempt to understand with a finite-size analysis whether $\Lambda_{\alpha\beta}$ has the necessary correlations with local structure to hope to extract elastic moduli from it. To this end, we once again consider how the variance of the strain tensor scales with the size of the coarse graining region used to construct it. In Fig.~\ref{fig:moduli_ldt_L} we see that the variance of $\Lambda_{xx}+\Lambda_{yy}$ (which one hopes to interpret as the bulk compliance) scales with $L^{-3}$ in two dimensions, again in contrast to the $L^{-d}$ scaling that any sensible definition of strain must have. We conclude that although $\Lambda_{\alpha\beta}$ has more correlation with structure in the cage regime than in the ballistic regime, it still cannot be used to give a well-defined elastic moduli. This scaling with $L$ highlights a danger of fitting limited ranges of data without systematically checking the dependence on the coarse graining scale. In Appendix C, we explore the unphysical behavior of the inferred elastic moduli if one attempts to artificially fit small portions of the strain distributions to exponential decays.

\section{Discussion\label{sec:disc}}

With a combination of simulation and experimental data, we have demonstrated that for model disordered solids it is incorrect to connect exponential fits of $P(\Lambda_{\alpha\beta}^2)$ with elastic moduli. In the limit of small $\Delta t$, we have presented an analytic model for the distributions of squared strain deformation tensor components. The model makes clear that the curvature of $P(\Lambda_{\alpha\beta}^2)$ on a log-linear plot can be completely explained as a $\chi^2$ distribution coming from the square of a single underlying Gaussian distribution, i.e., rather than from heterogeneous distribution of local elastic moduli.

Furthermore, we have shown that for any choice of $\Delta t$ the ratio of variances of $\Lambda_{xx}$ and $\Lambda_{xy}$ for these systems is essentially constant, with the ratio depending on the $\Delta t$ window chosen but independent of the pressure of the sphere packings. In contrast, the global measurement of $G/B$ for these systems scales with the pressure of the packings, $G/B \sim \sqrt{p}$. Thus, the ratio of the variances of these distributions does not correctly capture $G/B$, i.e., contrary to assumptions made in the literature~\cite{Schall2007,Schall2014}.  Moreover, we find that the variance of the strain in the cage regime scales with $L^{-3}$ in two dimensions in both simulation and experiment. This implies that the strain, as computed via the best-fit affine transformation, does not have enough information about local structure to reliably extract elastic moduli. We contrast this with two-dimensional crystalline systems~\cite{Zahn2003, Zhang2009} where a similar protocol reported variances that scaled with $L^{-2}$ and elastic moduli consistent with other measurements were successfully extracted.

Despite the fact that we find the variances of the best-fit affine strain distributions to be unrelated to the elastic moduli, we note that previous studies in amorphous materials have found spatial strain-strain correlations with a quadrupolar signature in the nonlinear response involving particle rearrangements \cite{Jensen2014}. This hallmark of continuum elasticity is characterized by a power-law decay in the strain correlations far from the rearrangement. In the linear response of quiescent systems, by contrast, we are not aware of any experiment or simulation of amorphous solids that shows that the best-fit affine strains arising from thermal motion have detectable power-law correlations, although such correlations must exist. Indeed, Rahmani et al. find that, in the absence of external strain, the strain-strain correlation function decays exponentially (with a length scale on the order of a single particle diameter) \cite{Schall2014}. 

The difference in this respect between crystalline and amorphous materials can be understood by considering the low-frequency excitations of the respective systems. In colloidal crystals (e.g., the hexagonal lattices studied in Refs. \cite{Zahn2003, Zhang2009}) the only vibrational modes are longitudinal and transverse sound modes that are (a) intimately related to the elasticity of the system and (b) spatially extended. We expect, via the equipartition theorem, that thermal fluctuations will populate these modes, leading to extended and strongly correlated strain profiles. By contrast, disordered solids have a large population of vibrational modes that are extended but disordered (e.g., in the boson peak). These modes span the system but have exponentially decaying local spatial correlations. Thermal population of these modes leads to strain profiles whose correlations are similarly exponentially localized; this is another argument for why measuring affine best-fit strains generated entirely by thermal fluctuations does not lead to meaningful information about the elasticity of disordered materials. In contrast, measuring these quantities in actively stressed or strained systems may still lead to meaningful information, as was discussed in a different context in Ref. \cite{Barrat2009} (see also below).

The disordered solids whose elasticity we are testing have explicit length scales, whose scaling for our model system goes as $l^*\sim p^{-1/2}$ and $l_c\sim p^{-1/4}$ \cite{Lerner2014,Silbert2005}, below which continuum elasticity fails to describe the response of these solids to imposed forces and deformations. We can estimate the magnitude of these length scales as ranging from $l^* \approx 7.1 \sigma$ and $L_T\approx 2.7 \sigma$ at $p=10^{-2}$ to $l^* \approx 71 \sigma$ and $L_T\approx 8.4 \sigma$ at $p=10^{-4}$. The range of coarse-graining length scales considered may be compared to the characteristic size of structural heterogeneities that lead to fluctuations in the local elastic moduli.  In a jammed solid, these heterogeneities are expected to be on the scale of $\ell_T$, or at worst $\ell^*$.  At the pressures studied, we have chosen coarse-graining scales that satisfy both $L \ll \ell^*$ and $L \gg \ell^*$. In both of these limits, comparing these lengths with the data in Fig. \ref{fig:moduli_ldt_L} suggests that the measurement of local affine strains is unable to detect the presence of these length scales, further undermining the claim that $\textrm{var}\left[\Lambda_{\alpha\beta}\right]$ is, on its own, intimately related to the local moduli.

We note that other methods for defining local strain fields and then connecting these to local elastic moduli have been proposed. Tsamados et al. \cite{Barrat2009} studied a linear strain tensor, $\epsilon^{lin}$, constructed from a continuous displacement field that was itself a coarse graining of local particle motion. In contrast to the methodology discussed here, Tsamados et al. also measured a local stress tensor, computed by a similar coarse graining, and defined the local moduli to be the constant relating these coarse-grained strains to coarse-grained stresses. Unfortunately for experimental measurements, the computation of the local stress fields requires knowledge of interparticle forces, which are typically difficult to identify in experimental systems. However, it might be interesting to see whether the thermal fluctuations of this, or other, definitions of local strain can be used in an argument in the spirit of Schall et al.\cite{Schall2007}.  

Another approach that has proven fruitful is to extract the particle-displacement covariance matrix from microscopic measurements. The bulk and shear moduli of the system can then be estimated from the inferred longitudinal and transverse speeds of sound in the material \cite{Still2014}. However, this method suffers from a few noted disadvantages.  Most significantly, a large amount of data is needed before the covariance matrix converges; this amount increases linearly with the number of particles in the system, $L^d$, where $L$ is the system length and $d$ the dimensionality~\cite{Chen2013}.  Second, disordered systems contain excess vibrational modes at low frequency that obscure the longitudinal and transverse acoustic branches of the phonon spectrum in systems that are too small; this effect scales as $1/L$. Together, these issues limit the utility of the covariance-matrix-approach to systems that are neither too large nor too small. This, then, explicitly limits the use of this tool when the distributions and spatial organization of elastic moduli are of interest.

In closing, our results highlight the subtlety of measuring the elasticity in soft disordered systems: a methodology that has been well-validated for two-dimensional crystalline systems fails spectacularly when applied to numerical simulations of disordered soft repulsive disks and laboratory experiments on colloidal packings. In light of this failure, we emphasize the critical importance of validating new methods of probing elastic constants by first testing them systematically against model systems whose properties are known by more conventional elasticity measurements.

\begin{acknowledgments}
This work was supported by the UPENN MRSEC under award NSF-DMR-1120901 (SSS and AJL), and the Advanced Materials Fellowship of the American Philosophical Society (DMS). Y.X., T.S., and A.G.Y. were supported by the National Science Foundation under Grants Nos. DMR12-05463, DMR-1305199, PENN MRSEC DMR11-20901, and NASA NNX08AO0G
\end{acknowledgments}

\appendix

\section{Defining local non-affinity\label{sec:formalism}}
In this appendix and the following we derive a simple statistical model that, for small $\Delta t$, almost completely captures the behavior of the components of the best-fit affine deformation tensors discussed above. For convenience, we first review an equivalent formulation of the $D^2_{min}$ language. In Appendix \ref{sec:statmodel} we will employ this language to make simple estimates of the variances of $\Lambda_{\alpha\beta}$.

\subsection{Operator expressions for non-affinity}
Here we closely follow the language of Ganguly et al. \cite{Ganguly2013}. In what follows roman indices will refer to particles and Greek indices to spatial coordinates. We begin by defining the initial position of particle $i$, $r_{i\mu}^0$. This could be the position of the particle at time $t-\Delta t$ as in the $D^2_{min}$ definition, or we could take it to be the inherent structure position of particle $i$ or its time-averaged position. Displacements from these initial positions will be written as by $u_{i\mu}(t) = r_{i\mu}(t)-r_{i\mu}^0$. For computing the local non-affinity for particle $i$ we additionally define displacements relative to that particle as $\vec{\Delta}_j(t) = \vec{u}_j(t)-\vec{u}_i(t)$. Note that it is common to choose a reference \emph{position} about which to define a local coarse graining volume, instead of a reference particle. In that case $r_{i\mu}^0$ simply sets the origin of the local coordinate system, which does not change between time $t-\Delta t$ and time $t$. In the following we will drop the explicit dependence on $t$ in our expressions when it is clear from context.

We next define an intensive measure of the local non-affinity of displacements relative to a reference particle in an analogous way to $D_{min}^2$:
\begin{equation}
\chi_i =\frac{1}{N} \min_\Lambda \left[ \sum_{\langle ij \rangle} (\vec{\Delta}_j-\Lambda(\vec{r}_j^0-\vec{r}_i^0))^2 \right],
\end{equation}
where the sum runs over all particles $j$ in the neighborhood considered, and $N$ is the number of particles in that neighborhood. Without the factor of $1/N$ and taking $\vec{r}_j^0$ to be the particle position at time $t-\Delta t$, the nonaffinity $\chi$ is exactly equal to the definition of $D^2_{min}$ in the main text. Independent of the presence of the prefactor $1/N$ the tensor $\Lambda_{\alpha\beta}$ here is identical to the best-fit affine deformation tensor defined in the introduction. Dividing out by the number of neighbors has been previously used to study thin films and pillars, where particles near the interface have many fewer neighbors than those in the center of the sample \cite{Shavit2014, Schoenholz2014}. Nevertheless, since below we will be exclusively interested in the distribution of the components $\Lambda_{\alpha\beta}$ the choice of an extensive or intensive definition of the total nonaffinity is irrelevant.

In order to express both $\chi$, and especially $\Lambda$, in a convenient operator form we define the following matrices. First, where $d$ is the spatial dimension, we write the $(1\times dN)$ matrix
\begin{equation}
\Delta = \left(\Delta_{11},\ldots, \Delta_{1d},\Delta_{21},\ldots,\Delta_{2d},\ldots,\Delta_{Nd}   \right),
\end{equation}
which compactly writes all of the relative displacements in a convenient order. Next we define the $(d^2\times 1)$ matrix
\begin{equation}
\lambda = \left( \Lambda_{11},\ldots, \Lambda_{1d},\ldots,\Lambda_{dd}   \right)^T.
\end{equation}
This simply unwraps the components of the best-fit $\Lambda$ affine-deformation tensor into a 1 dimensional array. Finally, we define the $(dN\times d^2)$ matrix
\begin{equation}
R_{j\alpha,\gamma \gamma'}=\delta_{\alpha \gamma} (r_{j \gamma'}^0-r_{i \gamma'}^0).
\end{equation}
This is a particularly convenient matrix with which to describe the initial relative positions of particles in the neighborhood of the reference particle.

We are now in a position to express the non-affinity in a very compact fashion. With the above definitions we have 
\begin{eqnarray}
\chi &=& \frac{1}{N}\textrm{min}_\Lambda \left[ \Delta-R\lambda \right]^2 \\
&=& \frac{1}{N} \textrm{min}_\Lambda \left[ \Delta^T\Delta - \Delta^T R \lambda-\lambda^T R^T \Delta + \lambda^TR^TR\lambda \right]\nonumber
\end{eqnarray}
Taking $d\chi /d\lambda$ and solving gives the minimizing affine deformation:
\begin{equation}
\lambda = \left(R^TR\right)^{-1}R^T\Delta \equiv Q\Delta.
\end{equation}
Given this minimizing $\lambda$, the non-affinity can be written as
\begin{eqnarray}
\chi &=&\frac{1}{N} \left( \Delta - R Q \Delta\right)^2 \\
&=&\frac{1}{N}\Delta^T \left[ 1 - 2 R \left( R^TR\right)^{-1} R^T + R  \left( R^TR\right)^{-1}R^T   \right] \Delta\nonumber\\
&\equiv&\frac{1}{N} \Delta^T P \Delta\nonumber.
\end{eqnarray}
The above expression defines a projection operator $P=1-RQ$ which projects components of $\Delta$ onto the space of non-affine deformations. 

\subsection{Specialization to two dimensions}
For concreteness, we explicitly write down an expression for the components of $\Lambda$ for a two-dimensional system. Taking a reference position $r_{i\mu}^0$ to set the origin of our local coordinate system, we have $R_{j\alpha,\gamma\gamma'} = \delta_{\alpha\gamma} r_{j\gamma'}^0$ for each particle $j$ in the coarse-graining area that we choose. The matrix $(R^T R)^{-1}$ then has a simple structure:
\begin{equation}
(R^T R)^{-1} = \left( \begin{array}{cccc} C & B & 0 &0 \\B&A&0&0\\0&0&C&B \\ 0&0&B&A \end{array}  \right),
\end{equation}
where
\begin{equation}
\label{eq:abcdef}
\begin{array}{ccc}
a= \sum_j (r_{jx}^0)^2; &b= \sum_j r_{jx}^0r_{jy}^0; & c=\sum_j (r_{jy}^0)^2\\
&&\\
A=\frac{a}{ac-b^2}; & B=  \frac{-b}{ac-b^2}  ;&C=\frac{c}{ac-b^2}.
\end{array}
\end{equation}

Thus, in two dimensions the operator $Q$ can be written as a combination of $d^2\times d$ blocks, each of which looks like
\begin{equation}
Q=(R^T R)^{-1}R^T = \left( \cdots \begin{array}{cc} Cr_{jx}^0+Br_{jy}^0 & 0\\B r_{jx}^0+A r_{jy}^0&0\\0& Cr_{jx}^0+Br_{jy}^0 \\ 0&B r_{jx}^0+A r_{jy}^0 \end{array}\cdots  \right),
\end{equation}
This lets us compactly write any component of $\Lambda$ using $\lambda = Q\Delta$, e.g.
\begin{equation}
\label{eq:lambdaxy}
\Lambda_{xy} = \sum_j \Delta_{jx}\left( Br_{jx}^0+Ar_{jy}^0\right).
\end{equation}
By writing the best-fit affine transformation tensor as a linear operator acting on the fluctuations it is already clear that one would not in general expect, e.g., an exponential distribution of $\Lambda_{xy}^2$ for short $\Delta t$. The relative displacements $\vec{\Delta}_{j}$ can be assumed to be normally distributed, after which the algebra of random variables suggests that $\Lambda_{xy}^2$ has a $\chi^2$ form. In the next section we show that a simple statistical model reproduces the distributions of $\Lambda_{xy}$ that we observe in our simulations.

\section{Statistical model\label{sec:statmodel}}
Here we show that in a disordered material we can use the algebra of random variables to accurately predict the distributions associated with $\Lambda_{\alpha\beta}$ at short $\Delta t$. As seen in Fig. \ref{fig:collapse_sdt}, and as could be anticipated from the functional form of Eq. \ref{eq:lambdaxy} in the absence of symmetry constraints and correlations, all of the components of $\Lambda$ have nearly identical distributions when averaged over our disordered systems. Our goal in this section will be to predict the variance of $\Lambda_{\alpha\beta}$ as a function of the typical scale of the fluctuations of $\Delta$ and the size of the coarse-graining volume. To do so, we start from a simple model for single-particle positional distributions and build up to the distribution of the best-fit affine deformation tensor.

For simplicity we focus on the two-dimensional case, and our dominant assumption will be a lack of structural order in the square coarse-graining cells. Hence, for a square coarse-graining square of side length $L=2R$ we take the $r_{j\alpha}^0$ to be uniformly distributed in $(-R,R)$, i.e. to have a probability distribution given by
\begin{equation}
P_{r_{j\alpha}^0}(x) = \left\{ 
\begin{array}{cc}
\frac{1}{2R} & |x|<R \\ 0 & |x|>R
\end{array}
\right. .
\end{equation}
The building blocks of the best-fit affine deformation tensor involve sums of products of these single-particle distributions. It is straightforward to show that 
\begin{equation}
P_{r_{jx}^0r_{jy}^0}(x) = \left\{ 
\begin{array}{cc}
\frac{1}{2R^2}\log\left(  \frac{R^2}{|x|}\right) & |x|<R^2 \\ 0 & |x|>R^2
\end{array}
\right. ,
\end{equation}
\begin{equation}
P_{(r_{j\alpha}^0)^2}(x) = \left\{ 
\begin{array}{cc}
\frac{1}{R\sqrt{x}} & 0<x<R^2 \\ 0 & \textrm{otherwise}
\end{array}
\right. .
\end{equation}

To make further progress we invoke the central limit theorem to describe the $a,b,$ and $c$ random variables. Let $n=\rho (2R)^d$, where $\rho$ is the number density, denote the average number of particles in a local coarse-graining volume, and the symbol $\mathcal{N}\left( \mu,\sigma\right)$ denote a Gaussian distribution with mean $\mu$ and width $\sigma$. Then we approximate
\begin{eqnarray}
a(x) = \sum_{j=1}^n P_{(r_{j\alpha}^0)^2}(x)  &\approx& \mathcal{N}\left(\frac{n R^2}{3},\frac{2R^2\sqrt{n}}{\sqrt{45}}\right) \approx c(x), \nonumber\\
b(x) =  \sum_{j=1}^n P_{r_{jx}^0r_{jy}^0}(x) &\approx & \mathcal{N}\left(0,\frac{R^2\sqrt{n}}{3}\right).
\end{eqnarray}

We next approximate the denominators that appear in the random variables $A, B,$ and $C$, i.e. $(ac-b^2)$. The $b^2$ part is trivial, and is given by
\begin{equation}
P_{b^2}(x)\approx \frac{3\exp \left( \frac{-9x}{2 n R^4} \right)}{\sqrt{2 \pi n R^4 x}}.
\end{equation}
The product $ac$ can be written as the sum of two general $\chi^2$ distributions:
\begin{eqnarray}\label{eq:acmb2}
ac &=& \frac{(a+c)^2}{4}-\frac{(a-c)^2}{4}\\
&=& \frac{1}{4} \left[ \mathcal{N}(\frac{2 n R^2}{3},\sqrt{\frac{8 n}{45}}R^2)  \right]^2 - \frac{1}{4} \left[ \mathcal{N}(0,\sqrt{\frac{8 n}{45}}R^2)  \right]^2\nonumber.
\end{eqnarray}
Note that both the second term in Eq. \ref{eq:acmb2} and the distribution of the $b^2$ have their weight centered about zero, whereas the first term in Eq. \ref{eq:acmb2} has a large positive mean. For simplicity we thus approximate the expression $(ac-b^2)$ by a single non-central $\chi^2$ random variable:
\begin{equation}
ac-b^2\approx \frac{1}{4} \left[ \mathcal{N}(\frac{2 n R^2}{3},\sqrt{\frac{8 n}{45}}R^2)  \right]^2
\end{equation}
Defining $\sigma_{ac} = R^2\sqrt{8n/45}$, the quantity $4(ac-b^2)/\sigma_{ac}^2$ is a non-central $\chi^2$ random variable with non-centrality parameter $\lambda=5n/2$ and number of summed normal variables $k=1$. This allows us to write the first moment and variance of $(ac-b^2)$ as
\begin{equation}
\langle ac-b^2 \rangle \approx \frac{n R^4(2+5 n)}{45},
\end{equation}
\begin{equation}
\textrm{var} \left[ ac-b^2 \right] \approx \frac{8 n^2(1+5 n)R^8}{2025}.
\end{equation}

We now approximate the mean and variance of, e.g., $A$ and $B$ by the lowest order terms in the Taylor expansion for the ratio of random variables, neglecting any covariance. That is, for random variables $X$ and $Y$ we approximate
\begin{eqnarray}
\left\langle \frac{X}{Y} \right\rangle &\approx &\frac{\langle X \rangle \langle Y \rangle^2}{\langle Y\rangle^3}+\frac{\langle X\rangle \textrm{var} [Y]}{\langle Y \rangle^3}+\cdots \\
\textrm{var}\left[\frac{X}{Y}\right] &\approx &\frac{\textrm{var}[X] \langle Y \rangle^2}{\langle Y\rangle^4}+\frac{\langle X\rangle^2 \  \textrm{var} [Y]}{\langle Y \rangle^4}+\cdots
\end{eqnarray}
We find
\begin{eqnarray}
\langle A \rangle &=& \frac{15}{(2+5n) R^2} + \frac{120 (1+5n)}{(2+5n)^3 R^2}\\
\textrm{var}[A] &=& \frac{180}{n(2+5n)^2 R^4} +\frac{1800(1+5n)}{(2+5n)^4 R^4}\\
\langle B \rangle& =& 0\\
\textrm{var}[B] &=& \frac{225}{n(2+5n)^2 R^4}
\end{eqnarray}

The penultimate step is to consider the variance of the products $\Delta_{jx}Ar_{jy}^0$ and $\Delta_{jx}Br_{jx}^0$. We again simply assume that the $r_{j\alpha}$ are uniformly distributed and that the $\Delta_{j\alpha}$ are normally distributed with zero mean and width $\sigma_\Delta$. Using the relation that the variance of a product of random variables $X_i$ is
\begin{equation}
\textrm{var}\left[ X_1\cdot X_2 \cdots X_n \right] = \prod_i \left( \textrm{var}\left[ X_i \right] + \langle X_i \rangle^2 \right) -\prod_i \langle X_i \rangle^2
\end{equation}
we have that
\begin{eqnarray}\label{eq:appvars}
\textrm{var}\left[\Delta_{jx}A r_{jy}^0  \right] &=& \frac{75 \sigma_\Delta^2}{(2+5n)^2 R^2} +\frac{60 \sigma_\Delta^2}{n(2+5n)^2 R^2}\nonumber\\
&& +\frac{1800\sigma_\Delta^2(1+5n)}{(2+5n)^4 R^2}+\cdots \\
\textrm{var}\left[\Delta_{jx}B r_{jx}^0  \right] &=& \frac{75 \sigma_\Delta^2}{n(2+5n)^2 R^2} +\cdots
\end{eqnarray}

We are finally in position to evaluate the variance of the components of the best-fit affine deformation tensor. Since
\begin{equation}
\Lambda_{xy} = \sum_j \Delta_{jx}\left( Br_{jx}^0+Ar_{jy}^0\right)
\end{equation}
we assume that the variance from each particle in the local coarse-graining volume contributes identically, and thus have
\begin{equation}
\textrm{var}[\Lambda_{xy}]  \approx n \left( \textrm{var}\left[\Delta_{jx}A r_{jy}^0  \right] + \textrm{var}\left[\Delta_{jx}B r_{jx}^0  \right]  \right),
\end{equation}
where the variances in this equation are given by Eq. \ref{eq:appvars}. A comparison between this equation and the simulation data (using the measured $\sigma_\Delta$) is shown in Fig. \ref{fig:collapse_sdt}, where it is seen to be an excellent estimate of the variance: with no adjustable parameters, and completely ignoring correlations from excluded volume (or any other source), this simple model describes the variance measured in the simulations to within $10\%$.

\section{System preparation\label{sec:simdetail}}
\subsection{Simulation details}
\begin{figure} 
\centering{
\includegraphics[width=0.8\linewidth]{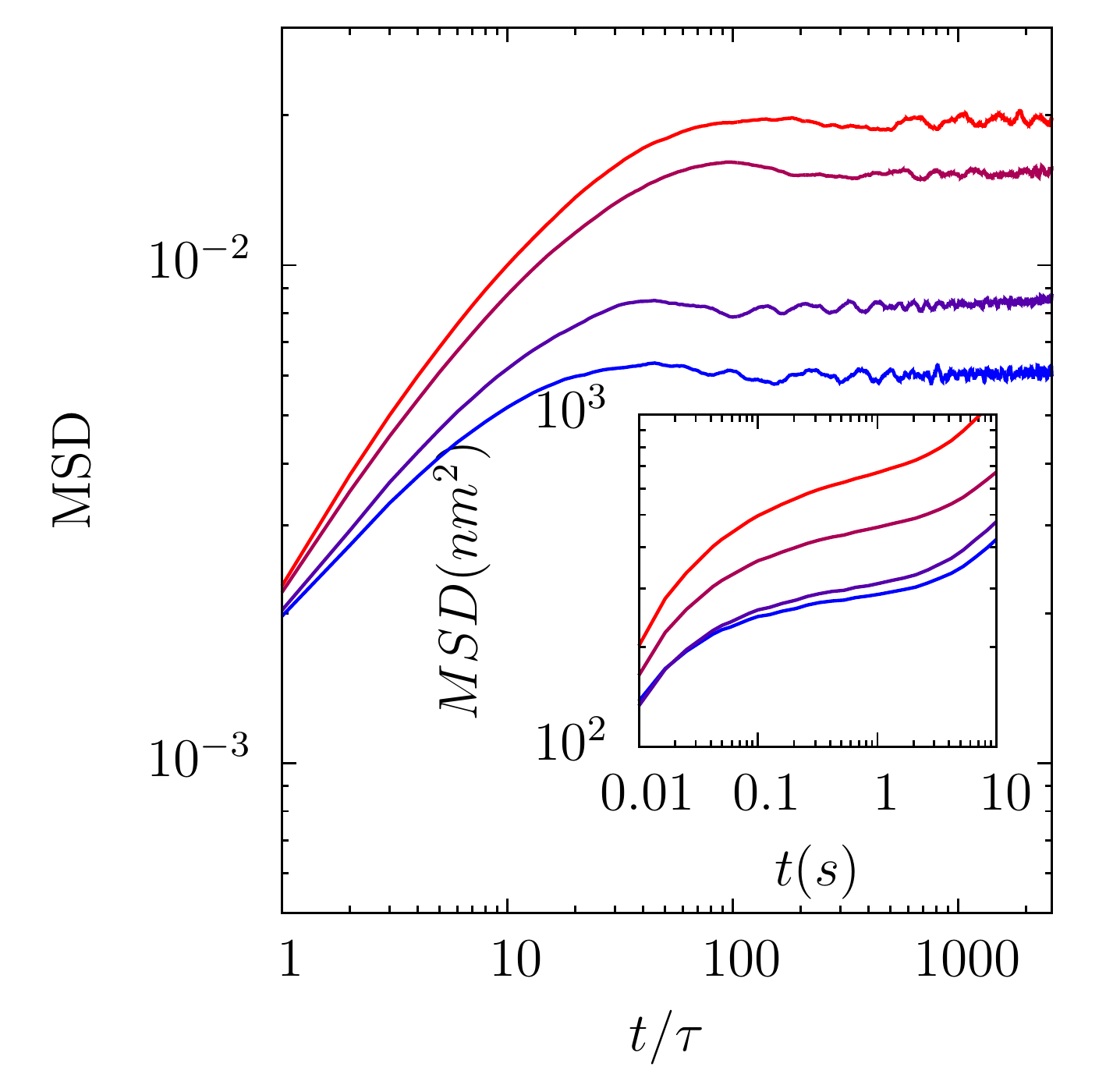} }
\caption{\label{fig:msd} (Color online) Mean-squared displacement (in units of particle diameters) for $p = 10^{-3.4},\ 10^{-3.0},\ 10^{-2.4},\ 10^{-2.0}$ (top to bottom). Inset. Mean-squared displacement in units of $nm$ for the colloidal system for $\phi = 0.8625, 0.865, 0.8695, 0.8775, 0.8822$ (top to bottom).
} 
\end{figure}

Our simulations are of frictionless packings with periodic boundary conditions composed of equal numbers of small and large spheres with a diameter ratio 1:1.4 and of equal mass, $m$. The particles interact with a repulsive, finite-ranged potential
\begin{equation}
V(r_{ij})=\left\{ \begin{array}{cc} 
\frac{\epsilon}{2}\left( 1-\frac{r_{ij}}{\sigma_{ij}} \right)^2 & r_{ij}<\sigma_{ij} \\
0 & r_{ij}>\sigma_{ij}, \end{array}
\right.
\end{equation}
where $r_{ij}$ is the distance between particles $i$ and $j$, $\sigma_{ij}$ is the sum of the particles' radii, and $\epsilon$ determines the strength of the interaction. We report energies in units of $\epsilon$ and distances in units of the average particle diameter. Time is measured in units of $\sqrt{\epsilon/(m \sigma^2)}$. We used this model to study 1024-particle systems in 2D and 4096 in 3D, for a range of pressures between $p=10^{-2}$ and $p=10^{-4}$. The initial configurations of these systems were set by first placing the particles at random in an infinite-temperature configuration, and then quenching to $T=0$ using a combination of linesearch methods, Newton's method, and the FIRE algorithm \cite{FIRE}. We then perform low-temperature molecular dynamics using the LAMMPS package \cite{LAMMPS}. For ease of comparing our timescales with the typical glassy crossover from ballistic to caged to diffusive behavior, mean-squared displacement curves for a subset of our 2D simulations are shown in Fig. \ref{fig:msd}.

In both the simulations and the experiments we compute the local non-affinity and best-fit affine deformation tensors as described in the text by partitioning the system into squares (cubes in 3D) of a given side-length. Thus, we compute with respect to a local origin of a coordinate system, rather than with respect to tagged reference particles. We have confirmed that this choice does not affect our conclusions.

\subsection{Experimental details}
Our experimental systems are quasi-two-dimensional packings of poly(N-isopropyl acrylamide) (PNIPAM) microgel particles. The full details of the experimental setup and data acquisition is reported in Ref. \cite{Still2013}. In brief, the disordered packing was prepared using a binary particle suspension with PNIPAM particles of two diameters: $\sigma_{1}\approx1.0\ \mu\textrm{m}$ and  $\sigma_{2}\approx1.4\ \mu\textrm{m}$. The sample was confined between two cover slips (Fischer Scientific) and then sealed from the edges with optical glue (Norland 63) \cite{Yodh2008}. Since PNIPAM is a temperature-sensitive polymer, the particle diameters can be controlled by changing the temperature. Thus, we tuned the effective packing fraction of the sample, $\phi$, \emph{in situ} using an objective heater (BiOptics). The temperature was set to a narrow range of $26.4 - 27.2^\circ$C so that the packing was above the jamming point. For each temperature studied, the trajectories of $N\approx 4500$ particles in the field of view were extracted from a total of $30,000$ frames of video using standard centroid-finding and particle-tracking techniques \cite{Crocker1996}. $D^2_{min}$ calculations were done by first subtracting the global drift of the sample and then using $\Delta t =0.273$ s, a value which is well into the plateau region of the mean-squred displacement \cite{Still2014}, as seen in the inset to Fig. \ref{fig:msd}.

\begin{figure} 
\centering{
\includegraphics[width=0.8\linewidth]{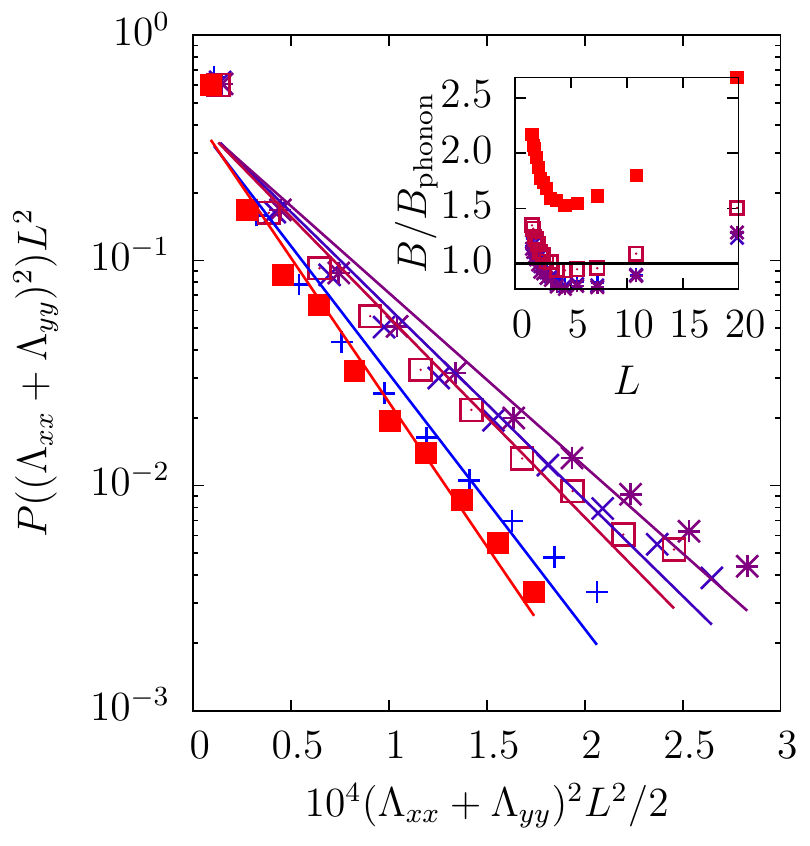} }
\caption{\label{fig:weird} (Color online) Normalized probability distributions of $(\Lambda_{xx}+\Lambda_{yy})^-2$ for the experimental colloidal system at a fixed volume fraction, truncated to a fixed dynamical range, for several different choices of the box coarse graining size. The fits are exponential decays, and the curves are calculated distributions for $L=1,3,4,11,20$ microns. Inset. Inferred modulus from the exponential decay fits normalized by the measured value by the methods in Ref. \cite{Still2014} as function of $L$. Data sets are for $\phi = 0.8625, 0.865, 0.8695, 0.8775, 0.8822$.
} 
\end{figure}

Under experimental conditions, where the full distribution and its heavy tail may not always be accessible, it may be tempting to try to fit to an exponential decay to these distributions over some limited dynamic range. Thus, in Fig. \ref{fig:weird} we briefly mention the results of approximating the observed experimental distributions of $(\Lambda_{xx}+\Lambda_{yy})^{-2}$ by exponential decays using an artificially restricted dynamic observation range. In the main plot we show exponential fits to the squared strain distributions as a function of coarse graining size, and there is little agreement between these fits. In the inset we plot the bulk modulus that would be inferred from such fits relative to the bulk modulus as measured in Ref. \cite{Still2014}. There is no systematic trend suggesting that, in the large $L$ limit where the method is ostesibly most sensible, the inferred modulus is asymptotically approaching the true value. This behavior could be anticipated from the results shown in Fig. \ref{fig:moduli_ldt_L}.

\bibliography{Strain_fluctuations_bib}

\end{document}